\def\aa{{A\&A}}
\def\aas{{A\&AS}}
\def\aj{{AJ}}

\def\apj{{ApJ}}
\def\apjs{{ApJS}}

\def\mnras{{MNRAS}}

\def\pasp{{PASP}}

\def\he4{{$^4$He}}
\def\Carb {{$^{12}$C}}
\def\Carbb{{$^{13}$C}}
\def\n14{{$^{14}$N}}
\def\Oxy{{$^{16}$O}}
\def\suns{{M$_{\odot}$}}

\documentclass[cup5b]{caps}
\usepackage{graphicx}
\usepackage{amssymb}
\usepackage{ociwsymp4}   

\HeadText{R.B.C. Henry}  

\begin{document}

\pagenumbering{arabic}

\author[]{R.B.C. HENRY\\University of Oklahoma}

%
%

\chapter{Element Yields of Intermediate-Mass Stars}

\begin{abstract}

Intermediate mass stars occupy the mass range between 0.8-8~\suns. In this contribution, evolutionary models of these stars from numerous sources are compared in terms of their input physics and predicted yields. In particular, the results of Renzini \& Voli, van~den~Hoek \& Groenewegen, and Marigo are discussed. Generally speaking, it is shown that yields of \he4, {\Carb}, and \n14 decrease with increasing metallicity, reduced mass loss rate, and increased rotation rate. Integrated yields and recently published chemical evolution model studies are used to assess the relative importance of intermediate mass and massive stars in terms of their contributions to universal element buildup. Intermediate mass stars appear to play a major role in the chemical evolution of \n14, a modest role in the case of {\Carb}, and a small role for \he4. Furthermore, the time delay in their release of nuclear products appears to play an important part in explaining the apparent bimodality in the distribution of damped Lyman-$\alpha$ systems in the N/$\alpha$--$\alpha$/H plane.

\end{abstract}

\section{The Nature of Intermediate-Mass Stars}

Intermediate-mass stars (IMS) comprise objects with ZAMS masses between 0.8 
and 8~\suns, corresponding to spectral types between G2 and B2.
The lower mass limit is the minimum
value required for double shell (H and He) fusion to 
occur, resulting in thermal pulsations during the asymptotic 
giant branch (AGB) phase and eventually planetary nebula formation.  Above the upper mass limit stars are capable of additional core-burning stages, and it is generally assumed that these stars become supernovae.
A Salpeter (1955) IMF 
can be used to show that IMS represent about 4\% of all stars above 
0.08~\suns, but this may be a lower limit if the IMF is flat at low stellar masses (Scalo 1998).

IMS evolution is an interesting and complex subject and the literature is extensive. A
good, complete, generally accessible review of the subject is given by 
Iben (1995). Shorter reviews focussing on the AGB stage can be found in Charbonnel (2002) and Lattanzio (2002). I will simply summarize here.

Intermediate-mass stars spend about 10-20\% of their nuclear lives in post main sequence stages (Schaller et al. 1992).
Fresh off the main sequence, a star's core is
replete with  H-burning products such as \he4 \& \n14. The shrinking core's temperature rises, a H-burning shell forms outward from the core, and shortly afterwards 
the base of the outer convective envelope moves inward and encounters 
these H-burning products which are then mixed outward into the envelope during 
what is called the {\it first dredge-up}. As a result, envelope levels of \he4, \n14, and 
{\Carbb } rise. Externally, the star is observed to be a red giant.

As the shrinking He core ignites, the star enters a relatively stable and quiescent time during which it synthesizes {\Carb } and \Oxy. Once core He is exhausted, the star enters the AGB phase, characterized by a CO core along with shells of H 
and He-fusing material above it.  Early in this phase, for masses in excess of 
4~\suns, {\it second dredge-up} occurs, during which the base of the convective envelope again extends inward, this time well into the intershell region,
and dredges up H-burning products, increasing 
the envelope inventory of \he4, \n14, and {\Carbb } as before.

Later in the AGB phase, however, the He shell becomes unstable to runaway fusion reactions, due to its thin nature and the extreme 
temperature sensitivity of He burning. The resulting 
He-shell flash drives an intershell convective pocket which mixes fresh {\Carb } outward toward the H-shell. But as the intershell expands, H-shell burning is momentarily 
quenched, and once again the outer convective envelope extends down into the intershell 
region and dredges up the fresh {\Carb } into the envelope, an event called {\it third dredge-up}. Subsequently, the intershell region 
contracts, the H shell reignites, and the cycle repeats during a succession of 
thermal pulses. Observational consequences of thermal pulsing and third dredge-up include the formation of 
carbon stars, Mira variables, and barium stars.

Now, in IMS more massive than about 3-4~\suns, the base of the convective 
envelope may reach temperatures which are high enough ($\sim$60 million K) to cause further H-burning via 
the CN cycle during third dredge-up. As a result, substantial amounts of {\Carb } are converted to \n14 in a 
process referred to as hot-bottom burning (Renzini \& Voli 1981; HBB). HBB 
not only produces large amounts of \n14 but also results in additional neutron production through the 
$^{13}$C($\alpha$,n)$^{16}$O reaction, where extra mixing is required to produce the necessary $^{13}$C. These additional neutrons spawn the production of 
s-process elements which are often observed in the atmospheres of AGB stars. Note that carbon star formation is precluded by HBB in those stars where it occurs. Other 
nuclei that are synthesized during thermal pulsing and HBB include $^{22}$Ne, 
$^{25}$Mg, $^{26}$Al, $^{23}$Na, and $^7$Li (Karakas \& Lattanzio 2003).

The thermal pulsing phase ends when the star loses most of its outer 
envelope through winds and planetary nebula (PN) formation, and thus the main fuel source for the H shell (and for the star) is removed and evolution is all but over. Note that the PN 
contains much of the new material synthesized and dredged up into the 
atmosphere of the progenitor star during its evolution. As this material 
becomes heated by photoionization, it produces numerous emission lines whose strengths 
can be measured and used to infer physical and chemical properties of the nebula.

\section{Stellar Models and Yields}

\subsection{The Predictions}

Models of intermediate mass star evolution are typically synthetic in 
nature. A coarse grid of models, in which values for variable quantities are computed directly 
from fundamental physics, is first produced. Then interpolation formulas 
are inferred from this grid which are subsequently used in a much larger run of models, thus reducing the computation time requirements. The 
models described below are of this type.

The major parameters which serve as input for IMS models include: stellar mass and 
metallicity, the value of the mixing length parameter, the minimum core mass 
required for HBB, the formulation for mass loss, and third dredge-up efficiency.

The first substantial study of IMS surface abundances using theoretical 
models was carried out by Iben \& Truran (1978), whose calculations accounted 
for three dredge-up stages including thermal pulsing. Renzini \& Voli (1981; RV) introduced hot bottom burning and the Reimers (1975) mass loss rate to their 
models and explicitly predicted PN composition and total 
stellar yields. van den Hoek \& Groenewegen (1997; HG) introduced a metallicity 
dependence, heretofore ignored, into their evolutionary algorithms along with 
an adjustment upwards in the mass loss rate,  the latter being a change driven 
by constraints imposed by the carbon star luminosity function (see below). 
Finally, Boothroyd \& Sackmann (1999) demonstrated effects of cool bottom 
processing on the {\Carb /\Carbb }  ratio; Marigo, Bressan, \& Chiosi (1996), Buell (1997), and Marigo (2001; M01) employed the mass loss 
formalism of Vassiliadis \& Wood (1993), which links the mass loss rate to the star's pulsation period, to predict yields of important CNO 
isotopes; and Langer et al. (1999) and Meynet \& Maeder (2002) studied the effects of 
stellar rotation on CNO yields. 

Table~\ref{t1} provides a representative sample of yield calculations carried out over the
past two decades.  To the right of the author column are columns which
indicate the lower and upper limits of the mass and metallicity ranges considered, an
indication of whether hot bottom burning or cold bottom processing was
included in the calculations (yes or no), the type of mass
loss used [R=Reimers (1975), VW=Vassiliadis \& Wood (1993)], an indication of whether the calculations included stellar
rotation (yes or no), and some important nuclei whose abundances were followed during the
calculations.

\begin{table}
\caption{Model Calculations of IMS Stellar Yields}
\label{t1}
\begin{tabular}{l|c|c|c|c|c}
\hline \hline
{Authors$^1$} & {M$_l$/M$_u$(\suns)} & {Z$_l$/Z$_h$} & {HBB/CBP} & {$\dot{M}$/Rot} & 
{Isotopes} \\
\hline
IT &1/8&.02&n/n&R/n&$^{12}$C,$^{13}$C,\n14,$^{22}$Ne \\
RV &1/8&.004/.02&y/n&R/n&$^{12}$C,$^{13}$C,\n14,$^{16}$O \\
HG &.8/8&.001/.04&y/n&R/n&$^{12}$C,$^{13}$C,\n14,$^{16}$O \\
FC &3/7&.005/.02&y/n&R/n& {\small $^7$Li,\Carb ,\n14, and more} \\
BS &.8/9&.0001/.02&n/y&R/n&$^{12}$C,$^{13}$C,\n14,$^{16,17,18}$O \\
BU & 1/8 & .002/.03 & y/n & VW/n & $^{12,13}$C,$^{14}$N,$^{16}$O \\
M01 &.8/6&.004/.019&y/n&VW/n&$^{12,13}$C,$^{14,15}$N,$^{16,17,18}$O \\
MM &2/7&10$^{-5}$&n/n&n/y&$^{12}$C,\n14,$^{16}$O \\
\hline \hline
\end{tabular}
$^1$References: IT=Iben \& Truran (1978); RV=Renzini \& Voli (1981); 
HG=v.d.Hoek \& Groenewegen (1997); FC=Forestini \& Charbonnel (1997); 
BS=Boothroyd \& Sackmann (1999); BU=Buell (1997); M01=Marigo (2001); MM=Meynet \& 
Maeder (2002) \\
\end{table}

It should be noted that the references listed in Table~\ref{t1} represent a much larger collection of papers which, while uncited here, nevertheless form an indispensible body of theory. The IMS enthusiast is hereby urged to consult the references in the table and explore this extensive literature.

For the purposes of a detailed discussion and comparison, I have singled out three yield sets in Table~\ref{t1} which are used most frequently to compute chemical evolution models.  These are the yields
published by RV, HG, and M01.
In the remainder of this subsection, each set will be considered by
itself, after which a comparison of their results will be made.

{\it Renzini \& Voli}: The calculations published by RV were the first to fully develop and include the process of hot bottom
burning.  Using parameterized equations in the spirit of Iben \& Truran
(1978), RV followed the surface evolution of $^{4}$He, $^{12}$C,
$^{13}$C, $^{14}$N, and $^{16}$O for stars of birth mass between 1 and
8~M$_{\odot}$ and metallicities 0.004 and 0.02.  They adopted the
empirical mass loss scheme of Reimers (1975) which at the time was 
uncalibrated
with respect to observables such as the carbon star luminosity function. The RV yields are still extensively used, although the models suffer from being unable to reproduce the carbon star luminosity function, probably because their adopted mass loss rate was too low (thus allowing more time for HBB to turn {\Carb } into \n14).

{\it van den Hoek \& Groenewegen}: The models of HG covered roughly the
same stellar mass range and isotope group as those of RV.  In addition,
HG used synthetic models with parameterized equations as did RV. HG also assumed a metallicity dependence of the algorithms, something RV had not done.
Although both RV and HG employed the Reimers mass
loss scheme, HG's $\eta$ parameter was over four times larger than
the one used by RV, thus elevating the mass loss rate and reducing stellar life times.  As we'll see below, this has a big effect
on the resulting yields.  HG also used
fundamental parameters for mass loss and dredge-up which were actually
tuned to reproduce observables such as the carbon star luminosity
function and the initial-final mass relation. 

{\it Marigo}: M01's calculations covered a stellar mass range 
between 0.8 and 5~\suns\footnote{Marigo considered effects of convective overshooting, which lowered the threshold for non-degenerate carbon burning as well as the upper mass limit for IMS.} and a metallicity range between 0.004 and 0.019.
Marigo's thermally pulsing AGB models were built upon the precursor models of Girardi
et al. (2000), which stopped at the onset of the thermally pulsing AGB
stage.  A major difference in Marigo's calculations compared with
earlier ones was the use of the mass loss scheme developed by
Vassiliadis \& Wood (1993) in which stellar pulsation theory and variable AGB star observations play roles. In addition, an apparent advantage of the Marigo yields for chemical evolution models is that Girardi-Marigo IMS models are based upon the same Padua library as are the massive star models and predicted yields of Portinari et al. (1997). Thus, the two sets can be combined to form a seamless yield set over a large stellar mass range (0.8-120~\suns).

\begin{figure}
\centering
\includegraphics[width=10cm,angle=270]{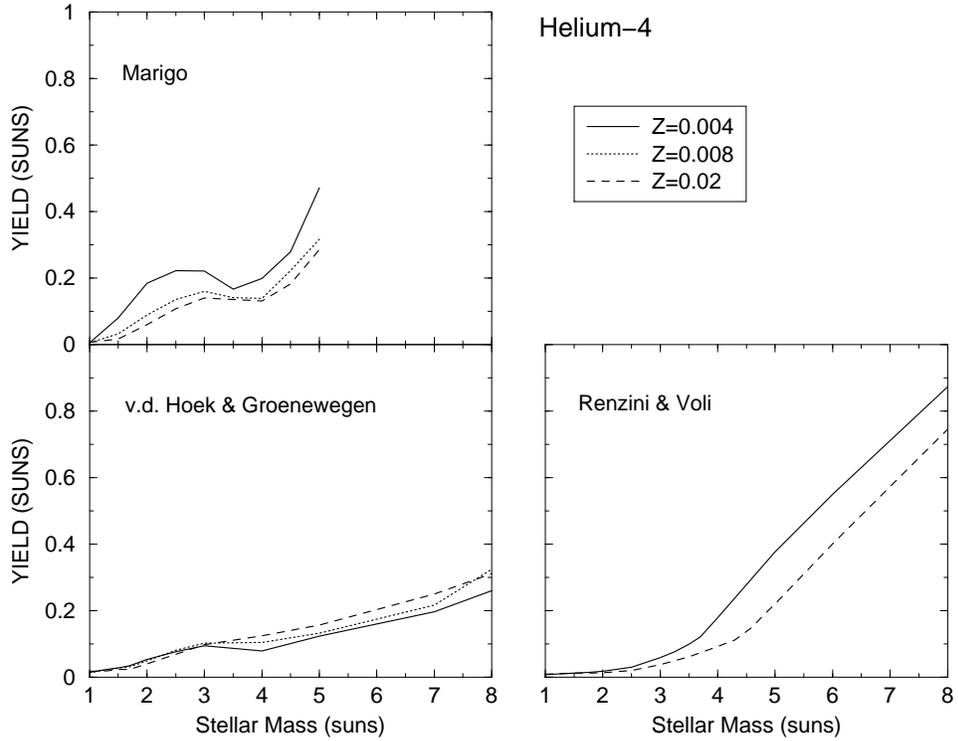}
\caption{Stellar yield for \he4 as a function of stellar mass, both in solar units, from M01 (top left panel), HG (bottom left panel), and RV (bottom right panel). Metallicity is shown in the legend and corresponds to line type in each figure.}
\label{phe4}
\end{figure}

\begin{figure}
\centering
\includegraphics[width=10cm,angle=270]{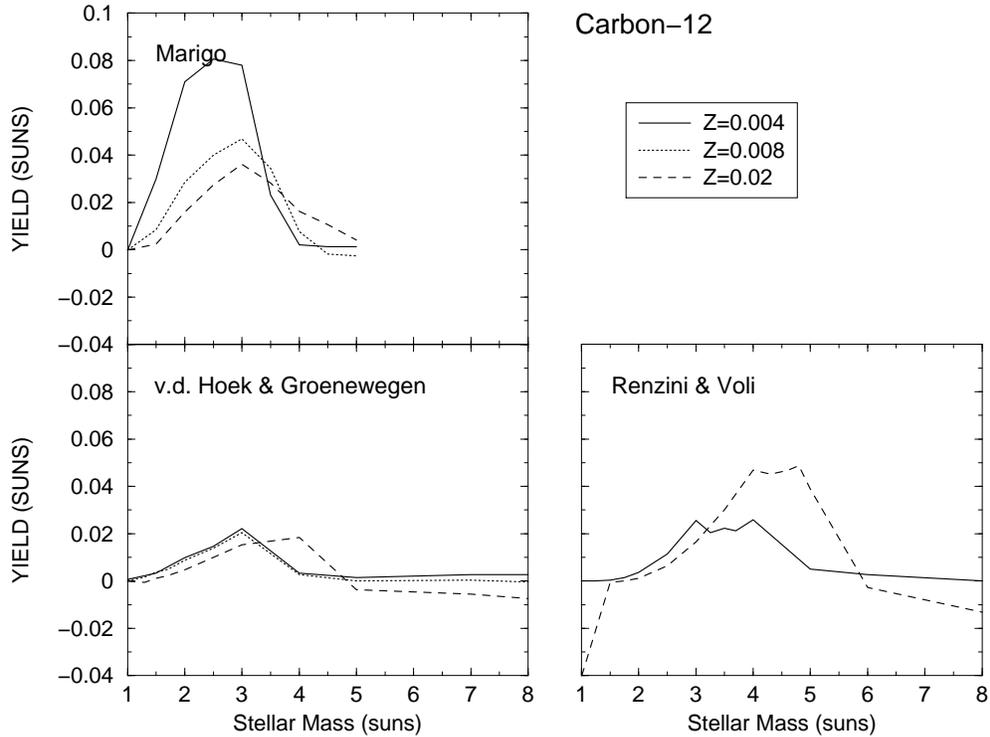}
\caption{Same as Fig.~\ref{phe4} but for $^{12}$C.}
\label{pc12}
\end{figure}

\begin{figure}
\centering
\includegraphics[width=10cm,angle=270]{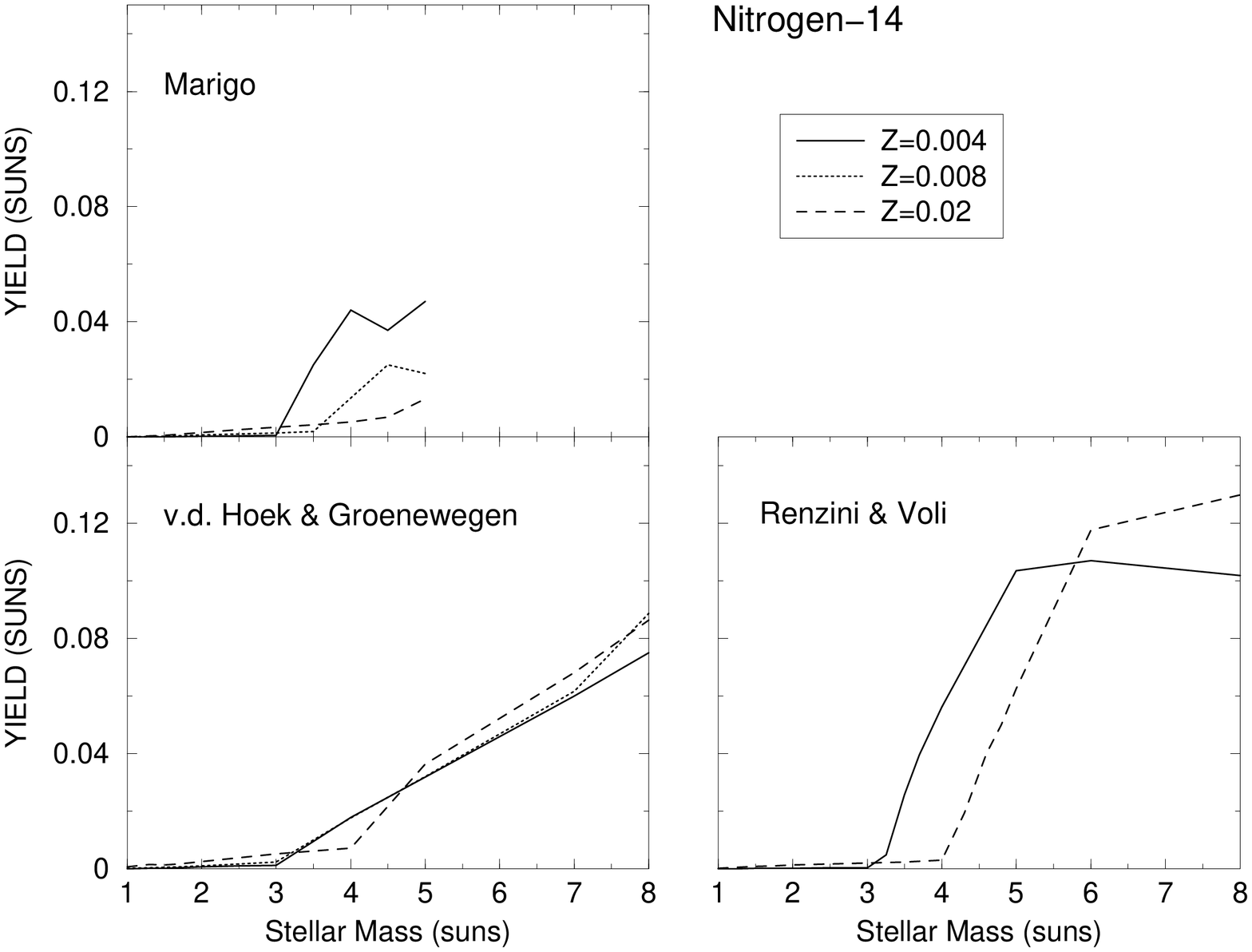}
\caption{Same as Fig.~\ref{phe4} but for $^{14}$N.}
\label{pn14}
\end{figure}

Now turning to the results of these three papers, figures \ref{phe4}, \ref{pc12}, and \ref{pn14} show comparisons of the yield 
predictions specifically for the isotopes of \he4, 
\Carb , and \n14, respectively. Because of space limitations, only these entities will be emphasized in the remainder of the paper.\footnote{IMS contributions to the production of many other isotopes, including those produced in the s-process, are discussed elsewhere in this conference by experts in the field. Interested readers should see the papers by Charbonnel, Karakas \& Lattanzio, Lambert, Busso, Straniero, and Reyniers \& Van Winckel for further updates and references.} Yields are plotted as a function of stellar 
birth mass, both in solar masses; metallicities are indicated by line type. 
Note that for ease of comparison, the three panels in each figure are identically scaled.

Generally speaking, IMS yields vary directly with the time between thermal pulses, the efficiency of third dredge-up, and the total number of thermal pulses which occur while the star is on the AGB. These factors are, in turn, fundamentally affected by stellar mass and metallicity as well as the values of the mixing length, dredge-up, and mass loss parameters. A full discussion of the relevant stellar theory is beyond the scope of this review, and the reader is referred to the papers listed in Table~\ref{t1} for more details and further references. Nevertheless, general statements can be made here concerning these relations. 

M01 points out that the number and duration of thermal pulses increases with stellar mass. Hence yields likewise tend to be related directly to this parameter, as can be seen clearly especially in the cases of \he4 and \n14 (Figs.~\ref{phe4} and \ref{pn14}), although the effect is overridden by HBB in the case of {\Carb } (Fig.~\ref{pc12}). In addition, the mass loss rate is directly related to metallicity, and so yields tend to be greater at low metallicity, a trend which is also visible in Figs.~\ref{phe4}, \ref{pc12}, and \ref{pn14}. The stellar core mass tends to be larger at lower metallicity (Groenewegen \& de~Jong 1993), and so the first thermal pulse dredges up more core material, adding to the other effects of metallicity. Also shown by M01 but not included here is the upward trend in yields with increasing values of the mixing length parameter (see Figs.~1 \& 2 in M01).

Many of the secondary features appearing in the figures relate to the details of HBB, which is effective in stars of mass greater than 3-4~\suns. Its efficiency increases directly with mass and inversely with metallicity (M01), and the effect is particularly apparent in the behavior of {\Carb } and \n14 yields in Figs.~\ref{pc12} and \ref{pn14}. For example, below the HBB mass threshold, the process does not operate, and the increase in the {\Carb } fraction in the envelope with third dredge-up is a major result of this process. However, above the threshold, {\Carb } is processed into \n14.

\begin{figure}
\centering
\includegraphics[width=10cm,angle=270]{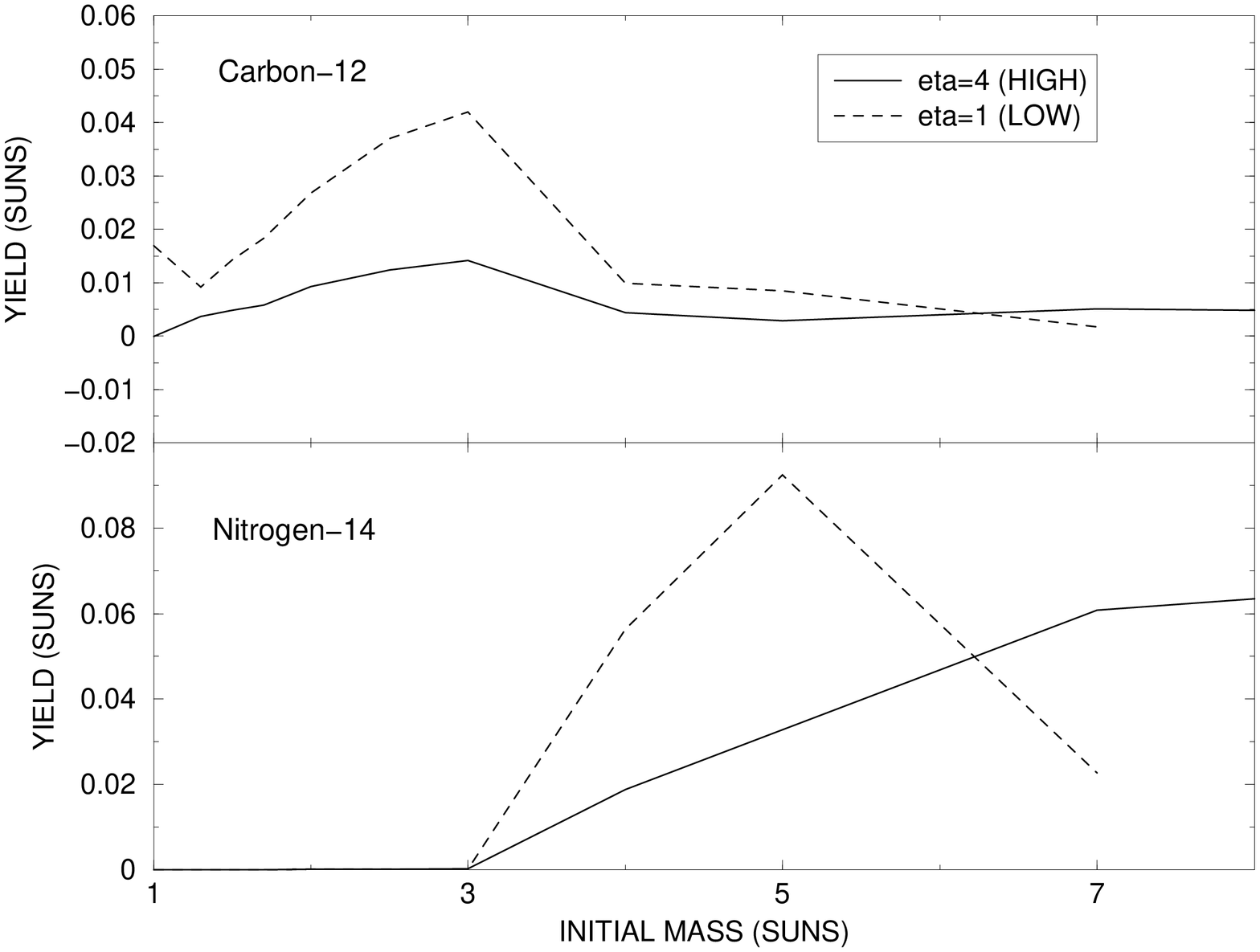}
\caption{Stellar yield versus initial stellar mass, both in solar units, for $^{12}$C and \n14, from HG. Results are shown for two different values of $\eta$, the mass loss parameter in the Reimers formulation.}
\label{massloss}
\end{figure}

The effect of mass loss rate is shown explicitly in Fig.~\ref{massloss}, which compares {\Carb } and \n14 yields predicted by HG, who employed the Reimers mass loss scheme, for two 
different values of the mass loss parameter. For $\eta$=1, i.e. relatively low 
mass low rate, yields in both cases are significantly higher than they are 
for the higher rate associated with the $\eta$=4 case. The relatively low mass loss parameter employed by RV explains why their \he4 and \n14 yields, in particular, tend to be greater than those of HG and M01.

\begin{figure}
\centering
\includegraphics[width=10cm,angle=270]{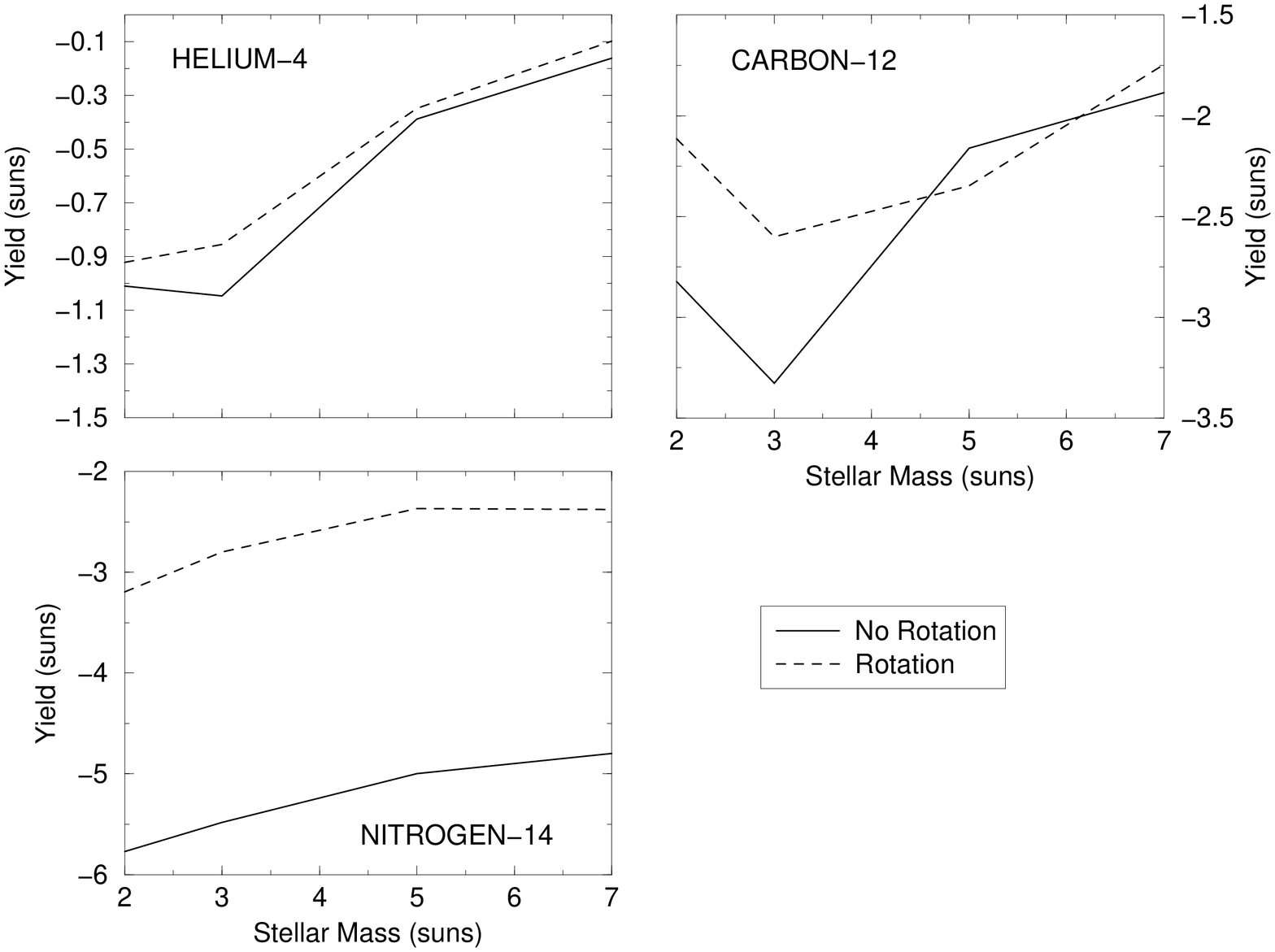}
\caption{Log of stellar yield versus stellar mass, both in solar units, for the situation of no rotation (solid line) and with rotation (v$_{ini}$=300 km/s; dashed line). Results are from Meynet \& Maeder (2002) for Z=10$^{-5}$.}
\label{rotation}
\end{figure}

The effects of stellar rotation on IMS evolution and yields have been receiving much attention lately. The topic is nicely explored by Langer et al. (1999),
Charbonnel \& Palacios (2003), and by Charbonnel (2003) at this conference. In terms of rotational effects on stellar yields, Meynet \& Maeder (2002) have recently computed an extensive model grid of stellar models for Z=10$^{-5}$ spanning the mass range of 2-60~\suns, and their results for the IMS mass range are shown in Fig.~\ref{rotation}
where the predicted yields for \he4, 
\Carb, and \n14 for the cases with and without rotation are compared\footnote{Meynet \& Maeder did not carry their IMS calculations past the first few thermal pulses, and therefore the effects of HBB are not included in their results.}. While rotation makes little 
difference in the case of \he4, it has a substantial effect on yields of {\Carb } 
(particularly at the low mass end) and \n14. According to Meynet \& Maeder, rotation increases the size of the CO core through more efficient rotational mixing; the effect is particularly pronounced at low Z, 
where the angular velocity gradient is much steeper.

Finally, Population~III nucleosynthesis for IMS has been studied by Chieffi et al. (2001), Marigo et al. (2001), and Siess, Livio, \& Lattanzio (2002). The Chieffi and Siess teams conclude that IMS are likely to be major contributors to {\Carb } and \n14 evolution in the early universe, although a great deal of work has yet to be done on this subject.

\subsection{Observational Constraints}

How well do the stellar models discussed above agree with observational 
constraints? At least four tests are available for comparison: the 
initial-final mass relation for IMS, the carbon star luminosity function, 
abundances in planetary nebulae, and {\Carb /\Carbb  } abundance ratios in red giants. 

First, realistic evolutionary models of IMS should predict the form of the initial-final mass 
relation that is observed. Typically, the empirical relation is established 
by comparing central star masses and turn-off masses in clusters (Weidemann 
1987; Herwig 1996; Jeffries 1997). M01 graphically compared these relations with theoretical results extracted from her models as well as those of HG and RV. The comparison makes clear that the higher mass loss rates used in the HG and M01 calculations are more successful in explaining the empirical relation than are the RV calculations with much lower rates.

Second, the paucity of high luminosity (high mass) carbon stars in the observed 
carbon star luminosity function (CSLF) compared with theoretical expectations, was first noted by Iben (1981). The 
finding gave rise to the introduction of HBB in order 
to reduce the amount of {\Carb } that is produced in the higher mass IMS. Thus, 
the CSLF provides another test for the models in general and mass loss in 
particular, because the latter, as we have seen, controls the number of pulses and resultant 
dredge-ups that occur on the AGB. Again M01 has compared model predictions with observations for the Magellanic Clouds and finds that the higher mass loss rates such as those used by HG and M01 are more appropriate than the lower ones employed by RV, where the latter predict too many high mass carbon stars.

\begin{figure}
\centering
\includegraphics[width=10cm,angle=270]{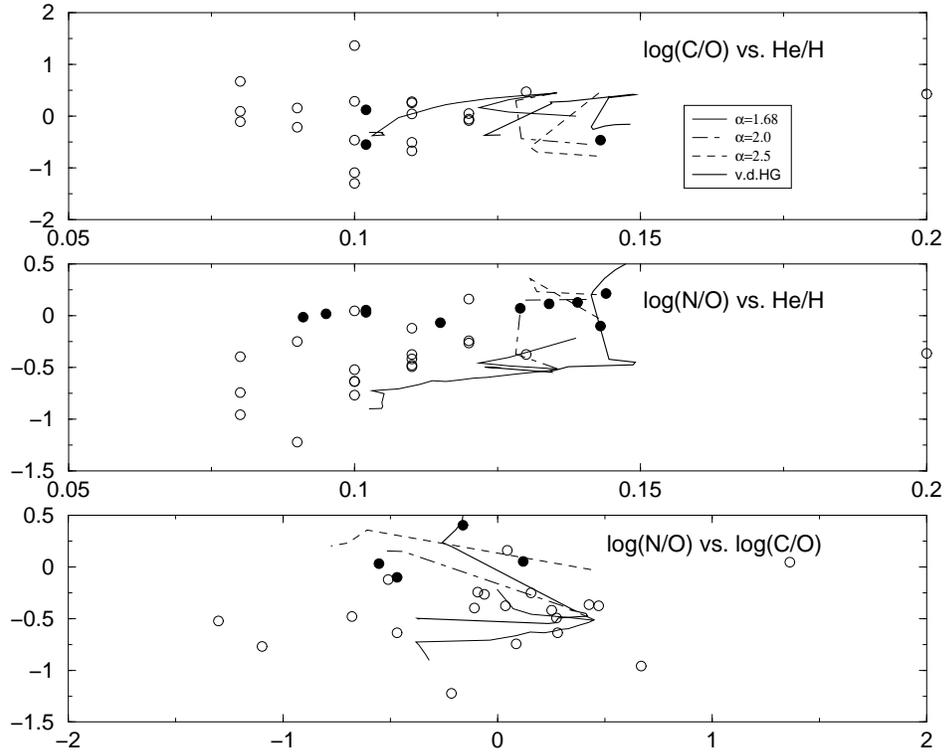}
\caption{{\it Top panel:} log(C/O) vs. He/H for a sample of planetary nebulae from Henry, Kwitter, \& Bates (2000; open circles) and Kingsburgh \& Barlow (1994; filled circles) compared with stellar evolution predictions by HG (bold line) and Marigo (2001; thin lines). Marigo's results are shown for three different values of the mixing length parameter, as indicated in the legend. {\it Middle panel:} Same as top panel but for log(N/O) vs. He/H. {\it Bottom panel:} Same as top panel but for log(N/O) vs. log(C/O).}
\label{hkb}
\end{figure}

Third, the stellar models also predict abundances in planetary nebulae (PNe) which 
can be compared directly with observations. Fig.~\ref{hkb} shows observed 
logarithmic N/O, C/O, and  normal He/H abundance ratios for a sample of PNe from Henry, 
Kwitter, \& Bates (2000; open circles) and Kingsburgh \& Barlow (1994; filled 
circles) along with model predictions by M01 (thin lines) for three different 
values of the mixing length parameter, as indicated in the legend. Also shown are model predictions from HG (bold lines). The models, all of which are for solar metallicity, 
occupy the same general area of the diagram as the observed points, and thus 
there appears to be consistency between theory and observation. Note 
that adjusting the mixing length parameter to higher values increases the 
extent of HBB and thus the amount of \he4 and \n14 predicted to be in the nebula (and in the yield). Kaler \& Jacoby (1990) studied the N/O ratio and central star 
mass for a sample of PNe and found that when central star progenitor masses exceeded about 3~\suns, the N/O ratios in the associated PNe were 
several times higher than in PNe with lower progenitor masses, a finding which suggests that HBB is effective in stars more massive than around 3~\suns. Recently, P{\'e}quignot et al. (2000) and Dinerstein et al. (2003) studied abundances of O, Ne, S, and Ar in a total of four low-mass metal-deficient PNe and found evidence for oxygen enrichment from third dredge-up in these objects. With a dozen or so Galactic halo and numerous Magellanic Cloud PNe now known, this implication of oxygen production by IMS needs to be investigated further.

Finally, the {\Carb /\Carbb  } ratio is predicted to be many times lower than its solar value in red giants due to the effects of first dredge-up (Charbonnel 1994). Recently, Smith et al. (2002) have reported new measurements of red giants in the LMC having [Fe/H] values between -1.1 and -0.3 which have {\Carb /\Carbb  } values consistent with expectations. Charbonnel (1994) presents models in which she predicts ratios in the observed range but claims that for stars of 2~{\suns} and below an additional mixing process (cool bottom processing) is required. More recent work by Boothroyd \& Sackmann (1999) supports this claim.

To summarize the picture for IMS stellar models and their predicted yields: 1. 
yields generally increase with stellar mass\footnote{The $^{12}$C yield increases with stellar mass up to the mass threshold for HBB, above which it declines.}, reduced mass loss rate, lower metallicity, and in the case of \n14, increased mixing length parameter; 2. Mass loss rates are important: models using rates consistent with observational constraints at the same time seem more likely to reproduce several other observed trends; 3. The 
threshold above which HBB occurs is somewhere between 3-4~\suns; and 4. Rotational mixing may substantially increase the yields of {\Carb } and primary \n14. Interesting areas which especially need to be developed further are the effects of rotation and zero metallicity (Pop.~III).

\section{Intermediate Mass Stars and Chemical Evolution}

In the grand scheme of galactic chemical evolution, do IMS matter? My discussion is divided into two parts. First, using computed values of integrated yields, I'll compare predicted IMS yields with those for massive stars. Then, the question of IMS evolutionary time delay and its effects on galactic systems will be addressed.

\subsection{Integrated Yields}

The relative impact of the two stellar groups, IMS and massive stars, can be assessed by integrating their yields over an initial mass function (IMF) for the two relevant mass ranges. Henry, Edmunds, \& K{\"o}ppen (2000) defined the integrated yield for isotope $x$, P$_x$, as:
\begin{equation}
P_x \equiv \displaystyle\int^{m_{up}}_{m_{down}}mp_x(m){\phi}(m)dm,
\end{equation}
where m$p_x$(m) is the stellar yield of isotope $x$ in solar masses of a star of mass m, $\phi$(m) is the IMF, and the upper and lower mass limits are m$_{up}$ and m$_{down}$, respectively. P$_x$ is then the mass of isotope $x$ which is newly produced and ejected per mass of new stars formed and ranging in mass m$_{down}$$\le$M$\le$m$_{up}$. For this specific exercise, IMS yields of RV, HG, and M01 along with massive star yields of Woosley \& Weaver (1995; WW) and Portinari, Chiosi, \& Bressan (1997; P) were used. Integrations were performed over a Salpeter (1955) IMF. 

Results of these calculations are provided in 
Table~\ref{t2}. The source of the yields is indicated in column~1, the lower and upper mass limits for the integration are given in columns 2 and 3, column~4 gives the metallicity applicable to the yields, and columns 5, 6, and 7 give the values of the integrated yield for \he4, {\Carb }, and \n14, respectively.

\begin{table}
\caption{Integrated Yields, P$_x$}
\label{t2}
\begin{tabular}{l|c|c|c|c|c|c}
\hline \hline
{Source$^1$} & {m$_{down}$} & {m$_{up}$} & {Z} & {P$_{^{4}He}$} & {P$_{^{12}C}$} & {P$_{^{14}N}$} \\
\hline
RV & 1 & 8 & 0.004 & 7.93E-3 & 6.05E-4 & 1.35E-3   \\
RV & 1 & 8 & 0.008 & 7.39E-3 & 2.14E-4 & 1.26E-3  \\
RV & 1 & 8 & 0.020 & 5.75E-3 & -9.57E-4 & 1.00 E-3  \\
HG & 1 & 8 & 0.004 & 6.14E-3 & 7.07E-4 & 5.70E-4  \\
HG & 1 & 8 & 0.008 & 6.41E-3 & 6.11E-4 & 6.24E-4 \\
HG & 1 & 8 & 0.020 & 6.41E-3 & 4.06E-4 & 7.29E-4  \\
M01 & 1 & 5 & 0.004 & 1.28E-2 & 3.74E-3 & 4.28E-4  \\
M01 & 1 & 5 & 0.008 & 7.47E-3 & 1.68E-3 & 1.89E-4  \\
M01 & 1 & 5 & 0.020 & 5.82E-3 & 1.09E-3 & 1.68E-4  \\
WW & 11 & 40 & 0.004 & 3.33E-2 & 9.92E-4 & 6.69E-5 \\ 
WW & 11 & 40 & 0.008 & 3.33E-2 & 9.92E-4 & 1.33E-4 \\
WW & 11 & 40 & 0.020 & 3.30E-2 & 9.94E-4 & 3.30E-4 \\
P & 6 & 120 & 0.004 & 4.78E-2 & 1.53E-3 & 1.72E-4 \\
P & 6 & 120 & 0.008 & 5.09E-2 & 6.56E-3 & 3.30E-4 \\
P & 6 & 120 & 0.020 & 6.28E-2 & 5.66E-3 & 8.09E-4 \\
\hline \hline
\end{tabular}
$^1$References: RV=Renzini \& Voli (1981); 
HG=v.d.Hoek \& Groenewegen (1997);  M01=Marigo (2001); WW=Woosley \& Weaver (1995); P=Portinari et al. (1998) \\
\end{table}

\begin{figure}
\centering
\includegraphics[width=10cm,angle=270]{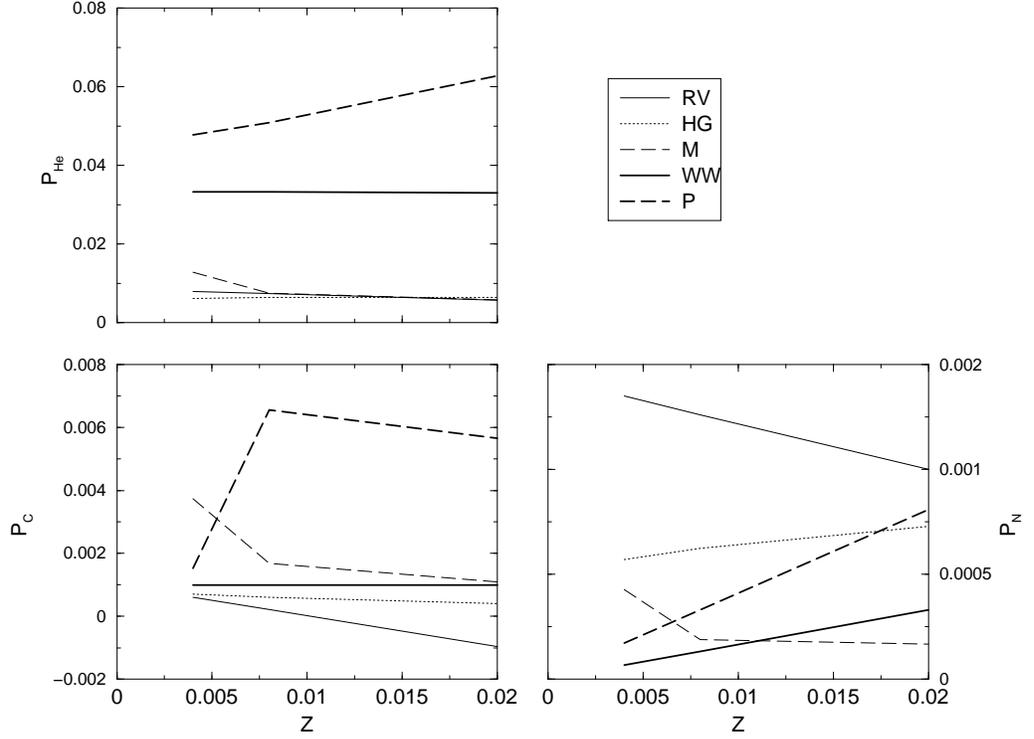}
\caption{Integrated yield, $P_x$, versus metallicity, Z, for \he4 (top left), {\Carb } (bottom left), and \n14 (bottom right). Plotted values are taken from Table~\ref{t2}.}
\label{p}
\end{figure}

Values for $P_x$ in Table~\ref{t2} are in turn plotted against metallicity in Figure~\ref{p}. The figure legend identifies the correspondence between line type and yield source, where the abbreviations are the same as those defined in the footnote to Table~\ref{t2}. Note that massive star integrated yields are indicated with bold lines, while thin lines signify IMS integrated yields.

For \he4 note that the three IMS yield sets predict similar results except at low metallicity, where the M01 yields are higher. According to M01, this difference is presumably due to the earlier activation and larger efficiency of third dredge-up in her models. It's also clear that IMS contribute to the cosmic buildup of \he4 at roughtly the 20-30\% level.

The RV yields for {\Carb } tend to be less than those of HG, while those of M01 are greater, due to differences in onset time and average efficiency of third dredge-up. Globally, the role of IMS in {\Carb } production is therefore ambiguous, because it depends upon which set of massive star yields one uses to compare with the IMS yields. For example, IMS yields are comparable to the massive star yields of WW, yet significantly less than those of Portinari et al.  

Finally, there is a significant difference between the three yields sets where \n14 is concerned. The lifetimes of the thermal pulses in the stars at the upper end of the mass range in M01's calculations are largely responsible for her \n14 yields being significantly less than the others'. On the other hand, RV's lower mass loss rate lengthens a star's lifetime on the late AGB and results in more \n14 production. When compared with massive star yields, RV and HG predict that IMS will produce several times more \n14, particularly at lower Z, when compared to either the WW or P yields. On the other hand, M01's models predict less \n14. 
Universally speaking, then, IMS yield predictions indicate that these stars contribute significantly to \n14 production, moderately to {\Carb }  productions, and hardly at all to \he4 production. Remember, though, that these conclusions are heavily based upon model predictions. The strength of these conclusions is only as strong as the models are realistic.

\subsection{IMS And Chemical Evolution Models}

The respective roles of IMS and massive stars in galactic chemical evolution can be further assessed by confronting observations of abundance gradients and element ratio plots with chemical evolution models which employ the various yields to make their predictions. Because there is a time delay of at least 30~Myr between birth and release of products by IMS, these roles may be especially noticeable in young systems whose ages are roughly comparable to such delay times or in systems which experienced a burst less that 30~Myr ago. 

Henry, Edmunds, \& K{\"o}ppen (2000; HEK) explored the C/O vs. O/H and N/O vs. O/H domains in great detail, using both analytical and numerical models to test the general trends observed in a large and diverse sample of galactic and extragalactic H~II regions located in numerous spiral and dwarf irregular galaxies. Using the IMS yields of HG and the massive star yields of Maeder (1992) they were able to explain the broad trends in the data, and in the end they concluded that while massive stars produce nearly all of the {\Carb } in the universe, IMS produce nearly all of the \n14. They also illustrated the impact of the star formation rate on the age-metallicity relation and the behavior of the N/O value as metallicity increases in low metallicity systems. In this conference, Moll{\'a}, Gavil{\'a}n, \& Buell (2003) report on their chemical evolution models which use the Buell (1997) IMS yields along with the WW massive star yields, where the former employ the mass loss rate scheme of Vassiliadis \& Wood (1993). Their model results confirm those of HEK in terms of the star formation rate and the age-metallicity relation.

Recently Pilyugin et al. (2003) reexamined the issue of the origin of nitrogen and found that presently the stellar mass range responsible for this element cannot be clearly identified because of limitations in the available data. 

\begin{figure}
\centering
\includegraphics[width=10cm]{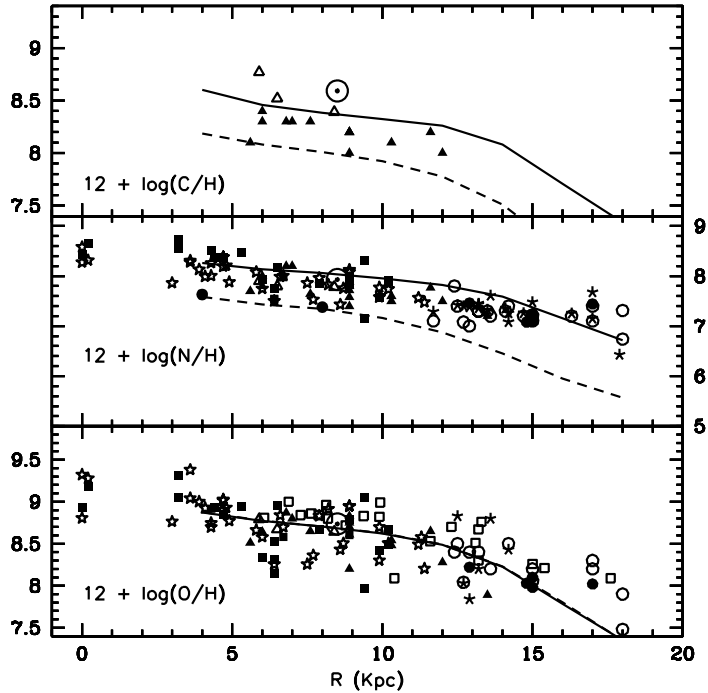}
\caption{12+log(X/H) vs. galactocentric distance, R, in kiloparsecs, as adapted from Fig.~10 in Chiappini et al. (2003). Data points show either H~II region or stellar abundances, where the references are detailed in their paper. The solid line is their model~7, while the dashed line has been added and is model~7 but with the IMS contributions to chemical evolution turned off.}
\label{chiappini}
\end{figure}

Chiappini et al. (2003) have explored the CNO question using chemical evolution models to study the distribution of elements in the Milky Way disk as well as the disk of M101 and dwarf irregular galaxies using the HG and WW yields. Like HEK, they conclude that \n14 is largely produced by IMS. However, they find that by assuming that the IMS mass loss rate varies directly with metallicity, {\Carb } production in these stars is relatively enhanced at low Z. In the end, they conclude that IMS, not massive stars, control the universal evolution of {\Carb}, in disagreement with HEK. 
Figure~\ref{chiappini} is similar to Fig.~10 in their paper and is shown here to graphically illustrate the effects of IMS on the chemical evolution of {\Carb } and \n14, according to their models. Each panel shows logarithmic abundance as a function of galactocentric distance in kpc for the Milky Way disk. Besides the data points, two model results are shown in each panel. The solid line in each case corresponds to the best-fit model in their paper, while the dashed line is for the same model but with the IMS contribution to nucleosynthesis turned off.\footnote{This altered model was kindly calculated by Cristina Chiappini upon the request of the author.} As can be seen, IMS make roughly a 0.5~dex and 1~dex difference in the case of C and N, respectively, i.e. their effects are sizeable.

Finally, the question of IMS production of nitrogen has become entangled in the debate over the interpretation of the apparent bimodal distribution of damped Lyman-$\alpha$ systems (DLAs) in the N/$\alpha$--$\alpha$/H plane (Prochaska et al. 2002; Centuri{\'o}n et al. 2003).\footnote{$\alpha$ represents elements such as O, Mg, Si, and S, whose abundances are assumed to scale in lockstep.} Most DLAs fall in the region of the ``primary plateau,'' located at a [N/$\alpha$] value of $\sim$-0.7 and between metallicities of -1.5 and -2.0 on the [$\alpha$/H] axis. However, a few objects are positioned noticeably below the plateau by roughly 0.8 dex in [N/$\alpha$], although still within the same metallicity range as the plateau objects. The Prochaska group proposes that these low N objects (LN-DLAs) correspond to systems characterized by a top-heavy initial mass function with a paucity of IMS, or, in the same spirit, a population of massive stars truncated below some threshold mass. Either possibility works through suppressing the IMS contribution to nitrogen production by reducing the proportion of these stars in a system's stellar population. The Centuri{\'o}n group, on the other hand, suggests that LN-DLAs are less evolved than the plateau objects, i.e. star formation occurred within them less than 30~Myr ago, so the LN-DLAs are momentarily pausing at the low-N region until their slowly evolving IMS begin to release their nitrogen. The latter picture, while not needing to invoke a non-standard IMF (an action which causes great discomfort among astronomers), does require that the time to evolve from the low-N ledge to the plateau region be very quick, otherwise their idea is inconsistent with the observed absence of a continuous trail of objects connecting these points. This problem is bound to be solved when the number of DLAs with measured nitrogen abundances increases, but it nevertheless illustrates an important role that IMS play in questions involving early chemical evolution in the universe.

\section{Summary}

Intermediate mass stars play an important role in the chemical evolution of \Carb, \Carbb, \n14, and $^{7}$Li as well as s-process isotopes. Stellar models have gained in sophistication over the past two decades, so that currently they include effects of three dredge-up stages, thermal pulsing and hot bottom burning on the AGB, metallicity, and mass loss by winds and sudden ejection. Generally speaking yield predictions from stellar evolution models indicate that yields increase as metallicity declines, as the mass loss rate is reduced, and when rotation is included. Furthermore, observational evidence supports the claim that the lower mass limit for hot bottom burning is between 3 and 4\suns.

Integration of yields over a Salpeter initial mass function shows clearly that IMS have little impact on the evolution of \he4 while at the same time playing a dominant role in the cosmic buildup of \n14. The case of {\Carb } is a bit more confused. The issue of \n14 production is particularly important in the current discussion of the distribution of damped Lyman-$\alpha$ systems in the N/$\alpha$--$\alpha$/H plane. 

Finally, what I believe is needed are grids of models which attempt to treat IMS and massive stars in a consistent and seamless manner. The role of each stellar mass range would be easier to judge if yield sets of separate origins did not have to be patched together in chemical evolution models. Otherwise, it is not clear to what extent the various assumptions which are adapted by stellar evolution theorists impact (and therefore confuse!) the analyses.

\section*{Acknowledgments}

I'd like to thank the Organizing Committee for inviting me to write this review and to present these ideas at the conference. I also want to thank Corinne Charbonnel, Georges Meynet, Francesca Matteucci, Cristina Chiappini, Jason Prochaska, John Cowan, and Paulo Molaro for clarifying my understanding on several topics addressed in this review. Finally, I am grateful to the NSF for supporting my work under grant AST 98-19123.

\end{document}